\documentclass[twocolumn,showpacs,preprintnumbers,superscriptaddress,amsmath,amssymb]{revtex4}

\usepackage{epsfig}
\usepackage{graphicx}
\usepackage{dcolumn}
\usepackage{bm}
\usepackage{times}
\usepackage{xcolor}
\usepackage[percent]{overpic}

\begin{document}


\title{Unification of theoretical approaches for epidemic
spreading on complex networks}

\author{Wei Wang} \email{wwzqbx@hotmail.com}
\affiliation{Web Sciences Center, University of Electronic
Science and Technology of China, Chengdu 610054, China}
\affiliation{Big Data Research Center, University of Electronic
Science and Technology of China, Chengdu 610054, China}
\affiliation{Center for Polymer Studies and Department of Physics,
Boston University, Boston, Massachusetts 02215, USA}

\author{Ming Tang} \email{tangminghuang521@hotmail.com}
\affiliation{Web Sciences Center, University of Electronic
Science and Technology of China, Chengdu 610054, China}
\affiliation{Big Data Research Center, University of Electronic
Science and Technology of China, Chengdu 610054, China}

\author{H. Eugene Stanley} \email{hes@bu.edu}
\affiliation{Center for Polymer Studies and Department of Physics,
Boston University, Boston, Massachusetts 02215, USA}

\author{Lidia A. Braunstein} \email{lidiabraunstein@gmail.com}
\affiliation{Center for Polymer Studies and Department of Physics,
Boston University, Boston, Massachusetts 02215, USA}
\affiliation{Instituto de Investigaciones F\'{i}sicas de Mar del
Plata (IFIMAR)-Departamento de F\'{i}sica,
Facultad de Ciencias Exactas y Naturales,
Universidad Nacional de Mar del Plata-CONICET,
Funes 3350, (7600) Mar del Plata, Argentina.}

\date{\today}
\begin{abstract}

\noindent
Models of epidemic spreading on complex networks have attracted great
attention among researchers in physics, mathematics, and epidemiology
due to their success in predicting and controlling scenarios of epidemic
spreading in real-world scenarios. To understand the interplay between
epidemic spreading and the topology of a contact network, several
outstanding theoretical approaches have been developed. An accurate
theoretical approach describing the spreading dynamics must take
both the network topology and dynamical correlations into
consideration at the expense of increasing the complexity of the
equations. In this short survey we unify the most widely used
theoretical approaches for epidemic spreading on complex networks in
terms of increasing complexity, including the mean-field, the
heterogeneous mean-field, the quench mean-field, dynamical
message-passing, link percolation, and pairwise approximation. We
build connections among these approaches to provide new insights into
developing an accurate theoretical approach to spreading dynamics on
complex networks.

\end{abstract}

\pacs{89.75.Hc, 87.19.X-, 87.23.Ge}

\maketitle

\section{Introduction} \label{sec:intro}

Throughout human history infectious diseases have been a constant threat
to the health of society, and when diseases become epidemic they cause
huge economic losses. Models that enable the prediction and control of
epidemics have attracted much attention among researchers in the fields
of epidemiology, biology, sociology, mathematics, and physics. Bernoulli
proposed the first mathematical model for understanding the spreading of
smallpox \cite{Bernoulli1760}. It initiated a new era in the modern
mathematical modeling of infectious diseases, and many models for
describing the characteristics of epidemic spreading in which the states
of individuals are disaggregated by compartments have been proposed. For
example, the acquired immune deficiency syndrome (AIDS) can be described
using the susceptible-infected (SI) model, since infected individuals,
once infected by the AIDS virus, cannot be cured. Seasonal influenza and
blennorrhagia can be described using the
susceptible-infected-susceptible (SIS) model, because individuals can be
infected more than once.  Chickenpox and measles can be modeled using
the susceptible-infected-resistant (SIR) model, because once infected
individuals have recovered they acquire permanent immunization
\cite{KEELING2007}.

During the last century many researchers assumed that all individuals
uniformly interact, i.e., that they interact with all other individuals
with the same probability \cite{Kermack1927, Anderson1992}. Thus the
internal structure or topology of the contact network through which the
epidemic spreads was neglected. The main features of the topology of a
contact network include its degree distribution $P(k)$, i.e., the
fraction of nodes with $k$ contacts, the weights on links and nodes,
degree correlations, clustering, and community structure
\cite{Newman2010a}.  With the availability of large-scale data in
real-world contact networks, scholars have become aware of the existence
of heterogeneities in the topology of networks (i.e., power-law degree
distribution and different weights of contact network), the high
clustering, and small-world phenomena
\cite{Albert2002,Newman2003,Holme2012, Boccaletti2014,Braunstein2003,
  Wu2006}. The effects of contact network topology on epidemic spreading
was first introduced by Pastor-Satorras and Vespignani
\cite{Pastor-Satorras2001} using complex networks in which the nodes
represent individuals and the edges the interactions among them.

Over the past decade this pioneering work has encouraged much
outstanding research in the field of network spreading dynamics
\cite{Pastor-Satorras2015a}.  Both Monte Carlo simulations
\cite{Ferreira2012,Shu2015,Fennell2016,Shu2016} and theoretical study
\cite{Pastor-Satorras2015} have investigated the effects of network
structures on epidemic spreading velocity
\cite{Barthelemy2004,Cui2014}, epidemic variability
\cite{Shu2012,Crepey2006}, epidemic size
\cite{Newman2001,Wang2014a,Yangrui2008,Wuqingchu2008,liu2016biologically,
wang2015immunity}, and epidemic
thresholds \cite{BOGUNA2013,CASTELLANO2010,Mieghem2013,
  Parshani2010,Lee2013,Wuqingchu2012}.  Both the epidemic size and
threshold can indicate the probability of an epidemic occurring
\cite{Parshani2010}, which seeds are influential
\cite{KITSAK2010,PEI2013,Zhong2015,Liu2015}, and how to effectively
control the epidemic once it begins \cite{PASTOR-SATORRAS2002a,
  COHEN2003,Yang2012a}.  When the transmission probability is above an
epidemic threshold the system is in an active epidemic state, i.e.,
there is a finite fraction of nodes infected by the epidemic, but when
the transmission probability is below the epidemic threshold the
epidemic dissipates.  Near the epidemic threshold the system exhibits
such interesting phenomena as the rare-region phenomenon
\cite{Goltsev2010,Odor2010, VOJTA2006,Buono2013} and the scaling
behavior of the main magnitudes \cite{Noh2009,Tome2005}.  Generally
speaking, previous research has addressed how the topology of the
contact networks has an affect on the macrocosmic, mesoscale, and
microscopic levels.  Research on the macroscopic scale, which focuses
primarily on the effects of degree and weight distributions
\cite{Pastor-Satorras2001,
  Pastor-Satorras2002,Yan2005,Yang2012,LinWang2014}, has revealed that
networks with strong heterogeneous degree distributions have a
vanishing epidemic threshold \cite{Pastor-Satorras2001} and that the
heterogeneous weight distributions increase the epidemic threshold
\cite{Kamp2013,Sun2014,Wang2014a}.  Research on the mesoscopic scale,
which focuses on degree-degree correlations, clustering, and
communities \cite{Moreno2003,WANG2013, MIN2013,Newman2009}, has found
that assortativity \cite{Moreno2003}, high clustering
\cite{Newman2009}, and community structure \cite{Liu2007} enhance the
epidemic outbreak and that disassortativity \cite{Moreno2003}
diminishes it. Research on epidemics from a microscopic point of view
\cite{Barthelemy2004,Barthelemy2005} has discovered that high-degree
nodes---hubs---are infected quickly \cite{Barthelemy2004} and that the
epidemic is more likely to transmit through low-weight edges
\cite{Xu2015,Buono2013}.

There are many successful theoretical approaches to describing epidemic
spreading on complex networks. When describing the interplay between
complex network structure and the dynamics of epidemic spreading, there
are two challenges.  The first is describing the intricate topologies of
the contact networks, since in real-world networks this involves
heterogeneous degree and weight distributions \cite{Newman2001b,
  Newman2001c}, high clustering \cite{Newman2001a,Newman2009,
  Clauset2009}, motif structure \cite{Milo2002,Cui2012}, community
structure \cite{Newman2006,Clauset2004}, and fractal structures
\cite{Song2009,Rozenfeld2007}. The second and more difficult challenge
is describing the strong dynamic correlations among the states of the
nodes. The dynamic correlations are produced when the epidemic being
transmitted to a node from two of its neighbors are correlated
\cite{Altarelli2014}. When there is only one seed node and there is high
clustering the dynamic correlations are obvious, since all the infection
paths come from the same seed. These two challenges are not fully
addressed in the existing literature, which always assumes (i) that an
epidemic spreads on a large, sparse network \cite{Moreno2002,
  Newman2001,Miller2012,Shrestha2015}, (ii) that dynamic correlations
among the neighbors do not exist \cite{Moreno2002}, and (iii) that all
the nodes or edges within a given class are statistically equivalent
\cite{Moreno2002,Wang2014}. These theoretical approaches can be changed
by removing or adding assumptions, but a comprehensive review of the
relationships among these approaches is still lacking. Here we discuss
the main contributions, basic assumptions, and derivation processes of
seven widely-used approaches in terms of their increasing complexity.

\section{Theoretical approaches} \label{sec:model}

\begin{figure}
\begin{center}
\epsfig{file=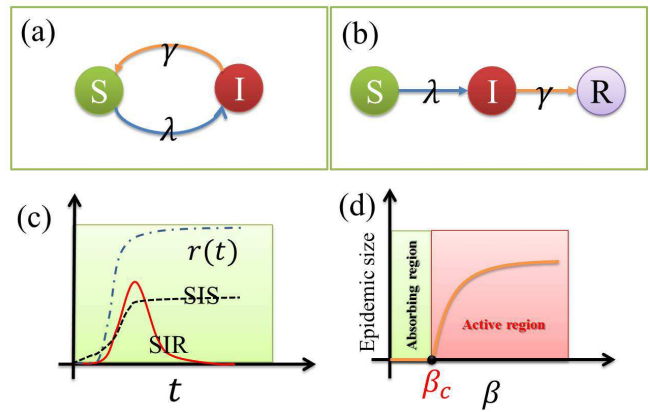,width=1\linewidth}
\caption{(Color online) Schematic representation of the
  susceptible-infected-susceptible (SIS) model (a),
  susceptible-infected-removed (SIR) model (b), the main magnitudes (c)
  and phase diagram the SIS and SIR models (d). In (b), the fraction of
  infected individuals $\rho(t)$ for the SIS model in the active state (
  black dashed line) and for the SIR model (solid red line), and the
  fraction of recovered nodes, $r(t)$ for the SIR model (dot dashed blue
  lines) in the active state. Notice that at the final of the epidemic
  $r(t)$ is constant.  In (d), the critical transmission rate (epidemic
  threshold) $\beta_c$ separates the plane into absorbing and active
  regions.  For $\beta\leq\beta_c$, there is no epidemic, i.e.,
  absorbing region; for $\beta>\beta_c$, the system has a global
  epidemic, i.e., active region.}
\label{fig1}
\end{center}
\end{figure}

The two most widely used model for epidemic spreading dynamics are the
reversible susceptible-infected-susceptible (SIS) and the irreversible
susceptible-infected-removed (SIR) models
\cite{Anderson1992,KEELING2007}. In the SIS model the nodes are either
susceptible or infected. In its continuous time version, at each time
step each infected node transmits its infection to all its susceptible
neighbors at the same rate $\lambda$ and returns to the susceptible
state at a rate $\gamma$. Thus the effective transmission rate, i.e.,
the effective transmission of infection, is
$\beta=\lambda/\gamma$. Without loss of generality we set $\gamma=1$. In
the SIR model, unlike the SIS model, an infected node recovers and is
permanently removed at the rate $\gamma$.  Figures~\ref{fig1}(a) and
\ref{fig1}(b) show schematically the transitions between compartments in
the SIS and SIR models. The schematic temporal evolutions of the SIS and
SIR models are shown in Fig.~\ref{fig1}(c).  When $t\rightarrow\infty$,
the order parameters (i.e., the epidemic sizes) of the two models
overcome a second-order phase transition depending on the value of
$\beta$, as shown in Fig.~\ref{fig1}(d). The critical threshold
$\beta_c$ divides the phase diagram into absorbing and active
regions. When $\beta\leq\beta_c$ there is an absorbing region and no
epidemic. When $\beta>\beta_c$ there is an active region and a global
epidemic develops.

The theoretical approaches that enable the computation of the epidemic
threshold and the magnitude of these models, i.e., the epidemic size,
exhibit similar but distinct frameworks for the SIS and SIR models and
are of two types.  In the first type the only difference between the
SIS and the SIR models is that in the SIR a removed state is added to
the SIS model. This first type includes the mean-field, heterogeneous
mean-field, dynamic message-passing, and pairwise approximation
approaches.  The second type includes the link percolation and
edge-based compartmental approaches, which provide the valid and
obvious framework for the SIR model since, unlike the SIS model, it is
irreversible.  Here we use the SIS model to illustrate the relations
among the existing approaches of this first type.  We use the SIR
model to explain the approaches of the second type. To clarify the
following, the definitions of the parameters are given in
Table~\ref{table1} in the Appendix.

\subsection{General frameworks for models of epidemic
  spreading}

{\bf Mean-field (MF) approach.}  The simplest framework for describing
these models---the mean-field (MF) approach---assumes that the
population is fully mixed, i.e., that all nodes in the population are
statistically equivalent and thus the interaction probabilities
between any two individuals are the same. As a consequence the
topology of the contact network is neglected. This approach was widely
used in the last century \cite{Anderson1992}. The MF approach also
assumes that there are no dynamic correlations among the states of a
node and its neighbors.  The time evolution of the density of infected
nodes in the SIS model is given by
\begin{equation} \label{men_field}
\frac{d\rho(t)}{dt}= -\rho(t)+\beta \langle k\rangle\rho(t)[1-\rho(t)],
\end{equation}
where $\rho(t)$ and $1-\rho(t)=s(t)$ are the fractions of infected and
susceptible nodes at time $t$, respectively, and $\langle k\rangle$ is
the average contact capacity of the nodes, i.e., the average degree of
the network. The first term on the right hand side of
Eq.~(\ref{men_field}) is the fraction of infected nodes that returns
to the susceptible, and the second term denotes the fraction of
susceptible nodes that are infected by infected neighbors. In the
steady state, i.e., $d\rho(t)/dt=0$, we have
\begin{equation} \label{final}
\rho(\infty)-\beta \langle k\rangle [1-\rho(\infty)]\rho(\infty)=0,
\end{equation}
where $\rho(\infty)= \rho(t \to\infty)$ is the fraction of infected
nodes in the steady state, i.e., the relative epidemic
size. Equation~(\ref{final}) has two roots with a trivial solution
$\rho(\infty)=0$, and a nontrivial solution $\rho(\infty) >0$ that
exists only when the effective transmission rate is greater than
an epidemic threshold
\begin{equation} \label{critical}
\beta_c^{{\rm MF}}=\frac{1}{\langle k\rangle}.
\end{equation}
The epidemic threshold $\beta_c^{\rm MF}$ divides the solutions into
absorbing and active regions. When $\beta\leq\beta_c^{\rm MF}$ there is
an absorbing region and no epidemic. When $\beta>\beta_c^{\rm MF}$ there
is an active region and a global epidemic develops. The simplest MF approach
neglects the internal structure of the contact networks and the
dynamic correlations among the states of the neighbors. This
oversimplified approach produces qualitatively analytical results,
such as the existence of an epidemic threshold and the scaling
relation of critical phenomena~\cite{Anderson1992}. For networks
with a homogeneous degree distribution (e.g., those with a
well-mixed population and those that are random regular
networks), the MF approach
accurately predicts epidemic size and threshold. Performing the
Taylor expansion of Eq.~(\ref{final}) at the epidemic threshold
$\beta=\beta_c^{\rm MF}$, Moreno et al. describe the epidemic
prevalence behavior as $\rho(\infty)\sim(\beta-\beta_c^{\rm MF})$
near the epidemic threshold~\cite{Moreno2002}.  Unfortunately the MF
approach can be inaccurate in some situations, e.g., when networks
have a heterogeneous degree distribution, since it neglects both
network topology and dynamical correlations. For example, Ferreira
et al.  numerically demonstrate that there is no epidemic threshold
for scale-free networks with a degree exponent $\nu\leq
3$~\cite{Ferreira2012}, but the MF approach predicts a finite value.

{\bf Heterogeneous mean-field (HMF) approach.}  To more accurately capture
network structure, Pastor-Satorras and Vespignani have improved the MF
approach for the SIS model by creating the heterogeneous mean-field (HMF)
approach \cite{Pastor-Satorras2001} in which nodes
with the same degree are equivalent. In the HMF approach the fraction of
nodes in the infected state $\rho(t)$ is split by the degree $k$ of the
nodes. Thus the primary magnitude is $\rho_k(t)$, which is the fraction
of infected nodes with degree $k$ at time $t$. The total fraction of
infected nodes is $\rho(t)=\sum_k P(k)\rho_k(t)$, where $P(k)$ is the
degree distribution of the network. In uncorrelated degree networks, a
susceptible node is connected to an infected neighbor with a probability
\begin{equation} \label{neighbor}
\Theta(t)=\frac{1}{\langle k\rangle}\sum_{k}^{k_{\rm max}} P(k)k\rho_k(t),
\end{equation}
where $k_{\rm max}$ is the maximum degree. The time
evolution of $\rho_k(t)$ is given by
\begin{equation} \label{h_men_field}
\frac{d\rho_k(t)}{dt}= -\rho_k(t)+\beta k [1-\rho_k(t)]\Theta(t).
\end{equation}
Similar to Eq.~(\ref{men_field}), the first (second) term in the right
hand side of Eq.~(\ref{h_men_field}) is the fraction of infected
(susceptible) nodes with degree $k$ that recover (are infected by
infected neighbors) at time $t$.

To obtain the epidemic threshold, Eq.~(\ref{h_men_field}) is linearized
around the initial conditions $\rho_k(0)\rightarrow0$ and then expressed
in a matrix form
\begin{equation}\label{HMF_J}
\frac{d\overrightarrow{\rho}(t)}{dt}= C
\overrightarrow{\rho}(t),
\end{equation}
where $\overrightarrow{\rho}(t)= [\rho_1(t),\cdots,\rho_{k_{\rm
      max}}(t)]^T$.  The Jacobian matrix
$C =\{ C _{kk^\prime}\}$ is given by
\begin{equation}\label{A_HMF}
 C _{k,k^\prime}=\beta\frac{kk^\prime P(k^\prime)}{\langle
k\rangle}-\delta_{k,k^\prime},
\end{equation}
where $\delta_{k,k^\prime}$ is a Dirac delta function.  The system has a
global epidemic---an active region---in which $\rho(t)=
\sum_{k}P(k)\rho_k(t)$ grows exponentially, which mathematically means
that the largest eigenvalue of $C$, $\beta\langle k^2\rangle/ \langle
k\rangle-1$, is greater than zero. Thus the epidemic threshold can be
expressed
\begin{equation} \label{h_thereshold}
\beta_c^{\rm HMF}=\frac{\langle k\rangle}{\langle k^2\rangle},
\end{equation}
where $\langle k\rangle$ and $\langle k^2\rangle$ are the first and
second moments of the degree distribution, respectively.  For
homogeneous networks, e.g., ER networks, the epidemic threshold is
$\beta_c^{\rm HMF}=1/(\langle k\rangle+1)$, which for small $\langle
k\rangle$ is different from that predicted using the MF approach [See
  Eq.~(\ref{critical})]. For heterogeneous networks with a power-law
degree distribution $P(k)\sim k^{-\nu}$, in the thermodynamic limit,
i.e., $N \to \infty$, the epidemic threshold is zero for degree
exponent $\nu\leq3$ due to the divergence of $\langle k^2\rangle$.
When $\nu>3$ there is a finite epidemic threshold. The HMF approach
has been highly successful in describing the dynamics of epidemic
spreading for two reasons, i.e., (i) we only need to know the degree
distribution and (ii) HMF is able to uncover how topological
heterogeneity affects epidemic spreading, e.g., there is no epidemic
threshold when the degree distributions are highly heterogeneous.
Researchers have generalized the HMF approach to investigate the
effects of weight distribution \cite{Chu2013}, degree-degree
correlations \cite{Boguna2003}, and multiplicity
\cite{Saumell-Mendiola2012,Wang2014,Liu2015a}. For example, Wang et
al. \cite{Wang2014,Liu2015a} generalized the HMF theory to study the
effect of asymmetrically-interacting spreading dynamics on complex
layered networks and found that the epidemic outbreak on the contact
layer can induce an outbreak of information on the communication
layer, and that the information spreading can effectively raise the
epidemic threshold.

The HMF approach is usually
effective when the networks have an infinite topological dimension,
i.e., when the number of nodes in a neighborhood grows
proportionately to network size $N$ with the topological distance
from an arbitrary origin~\cite{munoz2010griffiths}. Although random
networks above the percolation threshold indeed have an infinite
dimension, some researchers find that the HMF approach can fail
because two important factors are not taken into
consideration~\cite{durrett2010some,chatterjee2009contact,
givan2011predicting,Li2012a}. First, because the HMF approach
describes the network topology using degree distribution as the only
input parameter, the quenched connections among the nodes are
neglected.  Second, the dynamical correlations among the states of
neighbors are neglected, since Eq.~(\ref{h_men_field}) assumes that
the states of neighbors are independent~\cite{durrett2010some}. This
simplified assumption allows the HMF approach to accurately capture
spreading dynamics on annealed
networks~\cite{boguna2009langevin}. When epidemics spread on
quenched networks, the HMF description of the dynamics is only
qualitative~\cite{chatterjee2009contact}.

{\bf Quench mean-field (QMF) approach.}  Because neither the MF nor the HMF
approach can describe the full network structure, researchers use the
adjacency matrix $A$ to represent the full contact network topology. The
$A_{ij}$ component values of matrix $A$ are $A_{ij}=1$ when nodes $i$
and $j$ are connected. Incorporating the adjacency matrix, the quench
mean-field (QMF) approach is widely used to study the spreading
dynamics. Note that other approaches also use the adjacency matrix to
describe network topology, including the discrete-time Markov chain
approach \cite{Gomez2010} and the $N$-intertwined approach
\cite{Mieghem2011,Li2012}. At time $t$ a susceptible node $i$ is
infected by its neighbors with a probability $\beta\sum_{j=1}^N
A_{ij}\rho_j(t)$, where $\rho_j(t)=1-s_j(t)$ is the probability that
neighboring node $j$ of node $i$ is in the infected state at time $t$.
Thus the evolution of $\rho_i(t)$ can be expressed
\begin{equation} \label{q_men_field}
\frac{d\rho_i(t)}{dt}= -\rho_i(t)+\beta [1-\rho_i(t)]
\sum_{j=1}^NA_{ij}\rho_j(t)\;.
\end{equation}

The first (second) term on the right hand side of
Eq.~(\ref{q_men_field}) is the probability that node $i$ recovers
(becomes infected by its neighbors) at time $t$.  The fraction of nodes
in the infected state at time $t$ is $\rho(t)=1/N\sum_{i=1}^N\rho_i(t)$.
Since only a vanishing small fraction of nodes are in the infected state
at the beginning of the spreading, i.e., $\rho_i(0)\rightarrow 0$, we
linearize Eq.~(\ref{q_men_field}) around $\rho_i(t)\rightarrow 0$ and
rewrite it in matrix form
\begin{equation} \label{l_q_men_field}
\frac{d\overrightarrow{\rho}(t)}{dt}= -\overrightarrow{\rho}(t)
+\beta A\overrightarrow{\rho}(t),
\end{equation}
where $\rho_i(t)$ is element $i$ of the vector
$\overrightarrow{\rho}(t)=(\rho_1(t),\cdots,\rho_N(t))^T$.  Using the
same tool as that used to obtain Eq.~(\ref{h_thereshold}), the
epidemic threshold is given by
\begin{equation} \label{threshold_q}
\beta_c^{\rm QMF}=\frac{1}{\Lambda_A},
\end{equation}
where $\Lambda_A$ is the largest eigenvalue of $A$. The epidemic
threshold predicted by the QMF approach is dependent only on network
topology. In uncorrelated scale-free (SF) networks with a power-law
degree distribution, $\beta_c ^{\rm QMF}\propto \langle k\rangle/\langle
k^2\rangle$ when $\nu<2.5$, which produces the same threshold as
Eq.~(\ref{h_thereshold}). When $\nu>2.5$, $\beta_c^{\rm QMF}\propto
1/\sqrt{k_{\rm max}}$, which indicates that there is no epidemic
threshold in the thermodynamic limit \cite{Chung2003}. This result is in
contrast to the prediction from the HMF approach. The discrepancy
between the HMF and QMF approaches is addressed in
Refs.~\cite{Goltsev2010,Pastor-Satorras2015,BOGUNA2013,Lee2013}.

The QMF approach uses the
adjacency matrix to describe network topology and to predict
epidemic sizes and thresholds more accurately than those predicted
by the MF and HMF approaches.  However the dynamical correlations
among the states of neighbors are still neglected in
Eq.~(\ref{q_men_field}), and this produces deviations between the
theoretical predictions and numerical simulations.  Some researchers
doubt the accuracy of the epidemic threshold value predicted by the
QMF approach and some consider it completely wrong in some specific
situations~\cite{Lee2013,Goltsev2010,Ferreira2012,Shu2015}.  To
validate the effectiveness of the QMF approach in predicting a
epidemic threshold, Goltsev et al. \cite{Goltsev2010} define an
indicator, the inverse participation ratio (IPR), that quantifies the
eigenvector localization of $\Lambda_A$. The IPR of $\Lambda_A$ is given
by $v(\Lambda_A) =\sum_{i=1}^N f_i(\Lambda_A)^4$, where $f_i(\Lambda_A)$
is the $i$-th element of the eigenvector $\overrightarrow{f}(\Lambda_A)$
of $\Lambda_A$. If $\overrightarrow{f}(\Lambda_A)$ is delocalized,
$v(\Lambda_A) \propto O(0)$. If $\overrightarrow{f}(\Lambda_A)$ is
localized, $v(\Lambda_A)\propto O(1)$ \cite{Pastor-Satorras2015}.
Goltsev et al. \cite{Goltsev2010} claimed that
$\overrightarrow{f}(\Lambda_A)$ is delocalized for $\nu<2.5$, which
implies that the epidemic size is finite when $\beta>\beta_c^{\rm
  QMF}$. However $\overrightarrow{f}(\Lambda_A)$ is localized when
$\nu>2.5$, which means that only hubs and their neighbors are infected,
and as a consequence the epidemic grows very slowly and may die out due
to fluctuations. Thus they found that this localized state does not
constitute a true active state and that the epidemic threshold is closer
to that given by Eq.~(\ref{h_thereshold}).

Recently Pastor-Satorras and Castellano \cite{Pastor-Satorras2015a} have
further proven that $\overrightarrow{f}(\Lambda_A)$ of $\Lambda_A$ is
localized on the hubs when $\nu>5/2$ and localized on nodes with the
largest index in the \emph{K}-core decomposition \cite{KITSAK2010} when
$\nu<5/2$. To explain why the epidemic threshold predicted by the QMF
approach sometimes fails, we must understand the physical meaning of
$\overrightarrow{f}(\Lambda_A)$, which can be regarded the centrality of
a node (i.e., the eigenvector centrality) \cite{Newman2010a}.  The
eigenvector centrality assigns to each node a centrality proportional to
the sum of the eigenvector centralities of its neighbors. Unfortunately
the hubs with high centralities induce high centralities in their
neighbors, who in turn feed the centralities back to the hub.  As a
result, the centrality of the hub is overestimated.
Similarly we know that the
probability that a node is in the infected state is also
overestimated.  The variable $\rho_i(t)$ grows as $\rho_j(t)$
increases, and the value of $\rho_j(t)$ also increases when
$\rho_i(t)$ increases. Thus the infection is transmitted back and
forth through the same edge, which results in an ``echo chamber''
effect, and the infection probability of susceptible nodes is
overestimated~\cite{Radicchi2016}. We know that
Eq.~(\ref{q_men_field}) overestimates the probability that a node is
in the infected state.

\begin{figure}
\begin{center}
\epsfig{file=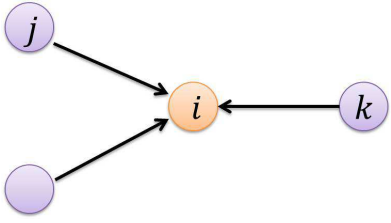,width=0.6\linewidth}
\caption{(Color online) An illustration of node in the cavity state. The
  test node $i$ is assumed to be in the cavity state, i.e. it can not
  transmit the decease to its neighbors but can be infected by them.
  The arrow direction indicates the direction of the infection.}
\label{fig3}
\end{center}
\end{figure}

{\bf Dynamical message passing (DMP) approach.}  To overcome the
weaknesses of the QMF approach but retain its advantages, i.e., to take
into consideration the full network structure, the dynamic
message-passing (DMP) approach was first proposed by Karrer and Newman
\cite{Karrer2010} in their study of the SIR model and generalized later
by Shrestha et al.  \cite{Shrestha2015} to describe the SIS model.  The
DMP approach disallows a node in the ``cavity'' state from transmitting
an infection to its neighbors but allows it to be infected by them (see
Fig.~\ref{fig3}). This prevents the epidemic from passing back and forth
through the same edge, causing an ``echo chamber'' \cite{Radicchi2016}
effect in Eq.~(\ref{q_men_field}) that decreases the overestimation of
the infection probability of susceptible nodes. Some dynamic
correlations among the states of the neighbors are also taken into
consideration.  The DMP approach is exact in tree-like networks
\cite{Karrer2010}, i.e., in networks with no loops.  Based on the DMP
approach, the time evolution of $\rho_i(t)$ can be written
\begin{equation} \label{dmp}
\frac{d\rho_i(t)}{dt}= -\rho_i(t)+\beta [1-\rho_i(t)]
\sum_{j=1}^NA_{ij}\theta_{j\rightarrow i}(t),
\end{equation}
where $\theta_{j\rightarrow i}(t)$ is the probability that node $j$ is
infected by its neighbors at time $t$ in the absence of node $i$
(i.e., node $i$ is in the cavity state).  The first term on the right
hand side of Eq.~(\ref{dmp}) takes into account the probability that
node $i$ recovers. The second term is the probability that it will be
infected by its neighbors.  If node $j$ recovers from the infected
state, $\theta_{j\rightarrow i}(t)$ will decrease, but if node $j$ is
infected by its neighbors in the absence of node $i$,
$\theta_{j\rightarrow i}(t)$ will increase with a
probability $\beta [1-\rho_j(t)] \sum_{\ell\in \mathcal{N}(j)\setminus
  i} \theta_{\ell\rightarrow j}(t)$, where $\mathcal{N}(j)$ is the set
of neighbors of node $j$.  Combining these two factors, the evolution
of $\theta_{j\rightarrow i} (t)$ can be written
\begin{equation} \label{dmp_edge}
\frac{d\theta_{j\rightarrow i}(t)}{dt}= -\theta_{j\rightarrow
  i}(t)+\beta [1-\rho_j(t)]
\sum_{\ell\in \mathcal{N}(j)\setminus i} \theta_{\ell\rightarrow j}(t).
\end{equation}

Using Eqs.~(\ref{dmp})--(\ref{dmp_edge}) we obtain the evolution of the
states of the nodes. Note that there will be $2E+N$ differential
equations, where $N$ and $E$ are the number of nodes and edges.
Initially $\theta_{j\rightarrow i}(0)\rightarrow 0$, since only a
vanishing small fraction of nodes are in the infected state.  Thus
linearizing Eq.~(\ref{dmp_edge}) around $\theta_{j\rightarrow i}(0)=0$,
Eq.~(\ref{dmp_edge}) can be rewritten
\begin{equation} \label{threshold_edge}
\frac{d\overrightarrow{\theta}(t)}{dt}=
\textbf{B}\overrightarrow{\theta}(t)-
\overrightarrow{\theta}(t),
\end{equation}
where $\textbf{B}$ is the non-backtracking matrix \cite{Martin2014}, and
$\theta_{j\rightarrow i}(t)$ is an element of the vector
$\overrightarrow{\theta}(t)$. The element of $\textbf{B}$ is
\begin{equation} \label{J_Matrix}
\textbf{B}_{j\rightarrow i,\ell\rightarrow h}=
\delta_{jh}(1-\delta_{i\ell}),
\end{equation}
where $\delta_{i\ell}$ is the Dirac delta function. The physical meaning
of $\textbf{B}_{j\rightarrow i,\ell\rightarrow h}$ is that when $i\neq
\ell$ the edge $\ell\rightarrow h$ can influence edge $j\rightarrow
i$. With arguments similar to those used to obtain
Eq.~(\ref{h_thereshold}), the epidemic threshold can be expressed
\begin{equation} \label{threshold_dmp}
\beta_c^{\rm DMP}=\frac{1}{\Lambda_\textbf{B}},
\end{equation}
where $\Lambda_\textbf{B}$ is the largest eigenvalue of the
non-backtracking matrix $\textbf{B}$.

The DMP approach is widely used
in such network science topics as spreading
dynamics~\cite{Karrer2010, Shrestha2015},
percolation~\cite{Karrer2014,hamilton2014tight,
radicchi2015breaking}, and cascading~\cite{radicchi2015percolation,
cellai2016message,son2012percolation}. It is widely applicable
because (i) it describes the complete network structure by using the
non-backtracking matrix, and (ii) it captures some dynamical
correlations among the states of neighbors by assuming that ``cavity''
nodes cannot transmit messages. This means that the DMP approach
produces exact results in networks that are tree-like.
Through extensive numerical simulations, Shrestha
et al.  \cite{Shrestha2015} found that the DMP approach accurately
predicts the SIS model in many
real-world networks. Similar results were found for the SIR model in
Refs.~\cite{Karra2010,Wang2015c}.

Note that the DMP approach has
    two drawbacks, (i) the equations are highly complex, and (ii) it is
    inaccurate in non-local tree-like networks. We recall that it is
difficult to analytically solve Eqs.~(\ref{dmp})--(\ref{dmp_edge})
because there are $2E+N$ differential
equations. To resolve drawback
    (i), we simplify the DMP approach by assuming that each edge has the
    same probability of connecting to infected neighbors. This
    simplified DMP (SDMP) approach and can be applied only to
    uncorrelated local tree-like networks (e.g., to uncorrelated
    configuration networks). Thus a susceptible node connects to an
infected neighbor with a probability
\begin{equation} \label{nei_inf}
\Theta(t)=\frac{1}{2E}\sum_{j\rightarrow i} \theta_{j\rightarrow i}(t).
\end{equation}
When we classify nodes according to their degree, for uncorrelated
networks Eq.~(\ref{nei_inf}) can be rewritten
\begin{equation} \label{nei_inf_un}
\Theta(t)=\frac{1}{\langle k\rangle}\sum_{k} (k-1)P(k)\rho_{k}(t),
\end{equation}
which was first derived by Barth\'{e}lemy and his collaborators
\cite{Barthelemy2004,Barthelemy2005}. Inserting Eq.~(\ref{nei_inf_un})
into Eq.~(\ref{h_men_field}), Barth\'{e}lemy et al.
predicted the velocity and hierarchical
structure of the epidemic spreading on scale-free networks. In this
approach the epidemic threshold is
\begin{equation} \label{hh_thereshold}
\beta_c^{\rm SDMP}=\frac{\langle k\rangle}{\langle k^2\rangle-
\langle k\rangle}.
\end{equation}

To resolve drawback (ii) we
decrease the ``echo chamber'' effect caused by finite
loops. Radicchi and Castellano introduce a more complicated DMP
approach to excluding redundant paths caused by triangles and
obtain more accurate predictions on both artificial and real-world
networks~\cite{Radicchi2016}. We still need to develop more accurate
approaches to describing the dynamics in real-world networks with
degree correlations, motifs, and community structures.

{\bf Pairwise approximation (PA) approach.}  The DMP approach cannot
accurately capture the dynamic correlations in non-tree-like
networks. The pairwise approximation (PA) approach best captures the
dynamic correlations \cite{Rand2009,Eames2008} by considering the
evolution of the pair node states, instead of the evolution of the
nodes. Denote $\psi_{x_i x_j}(t)$ as the probability that nodes $i$ and
$j$ are in the $x_i$ and $x_j$ states and $x\in\{S,I\}$, the following
relationships are fulfilled, $\psi_{I_iI_j}(t)+\psi_{S_iI_j}(t)
=\rho_j(t)$, $\psi_{I_iI_j}(t)+\psi_{I_iS_j}(t) =\rho_i(t)$,
$\psi_{S_iS_j}(t)+\psi_{S_iI_j}(t) =1-\rho_i(t)$, and
$\psi_{S_iS_j}(t)+\psi_{I_iS_j}(t) =1-\rho_j(t)$. With these equations
in mind, Eq.~(\ref{q_men_field}) can be written \cite{Mata2013}
\begin{equation}\label{pair}
\frac{d\rho_i(t)}{dt}= -\rho_i(t)+\beta
\sum_{j=0}^NA_{ij}\psi_{S_iI_j}(t).
\end{equation}
The first term is the probability that node $i$ recovers from the
infected state, and the second term is the probability that node $i$
becomes infected by its neighbors. If we neglect the dynamical
correlations between neighbors, i.e., $\psi_{S_iI_j}(t)=s_i(t)
\rho_j(t)$, Eq.~(\ref{pair}) reduces to Eq.~(\ref{q_men_field}).

In discussing the evolution of $\psi_{S_iI_j}(t)$, three events cause
$\psi_{S_iI_j}(t)$ to decrease, i.e., (i) node $j$ recovers from the
infected state, (ii) node $i$ is infected by neighboring node $j$ with a
probability $\beta\psi_{S_iI_j}(t)$, and (iii) the susceptible node $i$
is infected by another neighbor node $\ell$ with a probability
$\beta\sum_{\ell\in \mathcal{N}(i) \setminus j} \phi_{I_\ell
  S_iI_j}(t)$, where $\phi_{I_\ell S_iI_j}$ is the probability that node
$i$, $j$, and $\ell$ are respectively in the susceptible, infected, and
infected states at time $t$, and $\mathcal{N}(i)$ is the neighbor set of
node $i$. There are two events that cause $\psi_{S_iI_j}(t)$ to
increase, (i) node $i$ recovers from the infected state with a
probability $\psi_{I_iI_j}(t)$, and (ii) the susceptible node $j$ is
infected by another neighbor node $\ell$ with a probability
$\beta\sum_{\ell\in \mathcal{N}(j) \setminus i} \phi_{S_iS_jI_\ell}(t)$,
where $\phi_{S_iS_jI_\ell}$ is the probability that node $i$, $j$, and
$\ell$ are respectively in the susceptible, infected, and infected states
at time $t$, and $\mathcal{N}(j)$ is the neighbor set of node $j$. Based
on this, the evolution of $\psi_{S_i I_j}(t)$ is given by
\begin{equation}\label{pair_nei}
\begin{split}
\frac{d\psi_{S_i I_j}(t)}{dt}&=-\psi_{S_iI_j}(t)-\beta\psi_{S_iI_j}(t)
 -\beta\sum_{\ell\in \mathcal{N}(i) \setminus j}
\phi_{I_\ell S_iI_j}(t)\\
&+\psi_{I_iI_j}(t)+\beta\sum_{\ell\in \mathcal{N}(j) \setminus i}
\phi_{ S_iS_jI_\ell}(t).
\end{split}
\end{equation}
To complete Eq.~(\ref{pair_nei}), we apply a pair approximation,
i.e., we consider only the pair dynamic correlations as
\begin{equation} \label{close}
\phi_{x_i x_j x_\ell}(t)\approx\frac{\psi_{
x_i x_j}(t)\psi_{x_j x_l}(t)}{x_j (t)},
\end{equation}
where $x_j (t)$ is the probability that node $j$ is in the
$x\in\{S,I\}$ state at time $t$. Inserting Eq.~(\ref{close}) into
Eq.~(\ref{pair_nei}) and combining it with Eq.~(\ref{pair}), the
evolution of the the states of the SIS model can be described using
${N+E}$ differential equations.

To obtain the epidemic threshold, we linearize Eq.~(\ref{pair_nei})
around the initial conditions $\psi_{I_iI_j}(0)\rightarrow 0$ and
$\psi_{S_iS_j}(0)\rightarrow 1$. Using arguments similar to those for
obtaining Eq.~(\ref{h_thereshold}) we get the epidemic threshold when
the largest eigenvalue of $L$ is zero, where $L$ is the Jacobian matrix
of Eq.~(\ref{pair_nei}), and the elements of $L$ are \cite{Mata2013}
\begin{equation}\label{cri_pair}
L_{ij}=-(1+\frac{\beta_c^2k_i}{2\beta+2})\delta_{ij}
+\frac{\beta(2+\beta)}{2\beta+2}A_{ij}.
\end{equation}

At the expense of increasing the
complexity of the equations in the PA approach, we use the adjacency
matrix to accurately describe the full network topology, and we
capture the dynamical correlations among the states of neighbors by
considering the evolution of the pair node states. Performing
extensive simulations, Mata and Ferreira demonstrated that the
epidemic size and threshold predictions of the PA approach are more
accurate than those predicted using other methods, e.g., MF and
HMF~\cite{Mata2013}. Although the PA approach can capture some of
the dynamical correlations among the states of neighbors, solving the
above equations numerically is time-consuming, which hinders its wide
application.  In order to reduce the number of equations researchers
assume that all nodes of the same degree are statistically the
same~\cite{Szabo-Solticzky2016,
  Gross2009}. Thus we use
    $k_{\rm max}^2$ to describe the spreading dynamics because it
    decreases the complexity of the equations. Eames and Keeling, for
    example, used the PA approach to describe the spread of sexually
    transmitted diseases on heterogeneous networks, and their results
    are often in excellent agreement with simulations~\cite{ Eames2008}.
    Gross et al. used the PA approach to capture the assortative degree
    correlation, oscillations, hysteresis, and first order transition
    when an epidemic spreads on an adaptive
    network~\cite{Gross2007,Gross2009,bohme2011analytical}. Kiss et
    al. used the PA approach to theoretically predict non-Markovian
    epidemic spreading dynamics~\cite{Kiss2015}.  Recently researchers
    developed a generalized PA approach to study epidemic spreading on
    weighted complex networks~\cite{Rattana2013, Kamp2013}.

\subsection{Specific approaches using the SIR model}

{\bf Link percolation (LP) approach.}  In contrast to the reversible SIS model, the
irreversible SIR model allows us to examine the final state of the
epidemic at which an individual is either susceptible or recovered. The
most commonly-used approach is link percolation (LP) approach, and the most
studied version is the time-continuous Kermack-McKendrick \cite{Ker_01}
formulation in which an infected individual transmits the disease to a
susceptible neighbor at a rate $\lambda$ and recovers at a rate
$\gamma$. This SIR version has been widely studied in the
epidemiological literature, but unfortunately it allows some individuals
to recover immediately after being infected, which is unrealistic since
any real-world disease has a characteristic average recovery time. To
overcome this shortcoming, many studies use the discrete Reed-Frost
model \cite{Bai_01} in which an infected individual transmits the
disease to a susceptible neighbor with a probability $\lambda$ and
recovers $t_r$ steps following the time of infection. In the discrete
updating method the transmissibility $\beta$ is the probability that an
individual will infect one susceptible neighbor before recovery, and it
is given by
\begin{equation}
\beta=\sum_{u=1}^{t_r}\lambda(1-\lambda)^{u-1}=1-(1-\lambda)^{t_r}.
\label{Trans}
\end{equation}
Note that in the continuous time updating approach $\beta \approx
1-e^{-\lambda/\gamma} \approx \lambda/\gamma$ is used in
Ref.~\cite{Newman2001}.

\begin{figure}
\begin{center}
\epsfig{file=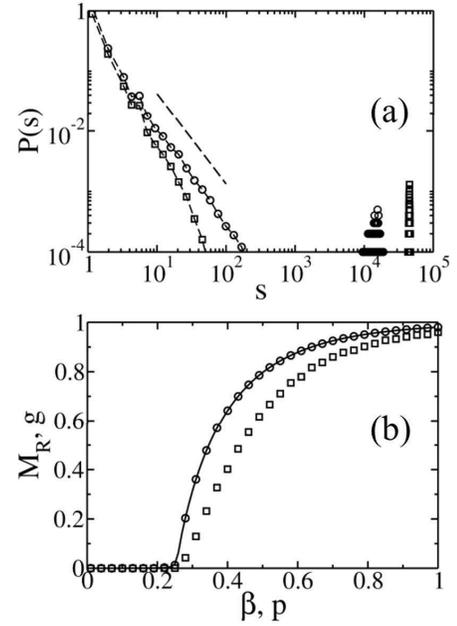,width=0.7\linewidth}
\caption{(Color online) Effects of the cutoff $s_c$ on the mapping
  between the SIR model and link percolation for a Poisson degree
  distribution network (ER) with $\langle k \rangle=4$
  ($\beta_{c}^{LP}=0.25$), $N=10^5$ . In (a) we show the probability
  $P(s)$ of a cluster of size $s$ (including the size of the giant
  component) in the SIR model for $\beta=0.27$ ($\bigcirc$) and
  $\beta=0.40$ ($\square$). We can see that the gap between the epidemic
  sizes and the distribution of outbreaks increases with $\beta$. As
  $\beta$ is at the threshold the slope is $\tau -1=3/2$ marked by
  dashed lines and $s_c\approx 200$.  In Fig. (b) we show the simulation
  results for $M_R$ for $s_c=1$ ($\square$) and $s_c=200$
  ($\bigcirc$). Note that when $s_c=200$, we average the final size of
  infected clusters only over epidemic realizations. Considering only
  the conditional averages, we can see that $M_R$ maps with $g$ (solid
  line) for $s_c=200$. The simulations are averaged over $10^4$
  realizations.}
\label{sc11}
\end{center}
\end{figure}

The order parameter $M_R(\beta) = M_R$, which is the final fraction of
recovered nodes, overcomes a second-order phase transition at the
epidemic threshold $\beta_c^{\rm LP}$, which is determined by the network
structure. Note that the Reed-Frost model can be mapped into a link
percolation process
\cite{Gra_01,Newman2001,Mil_01,Mey_01}. Heuristically, the relation
between SIR and link percolation is sustained because the probability
$\beta$ that a link is traversed by the disease is equivalent to the
occupancy probability $p$ in link percolation. Thus both processes have
the same threshold and belong to the same universality class. In
addition, each realization of the SIR model corresponds to a single
cluster of link percolation.  This feature is relevant when mapping
between the order parameters $g(p=\beta)= g$ of link percolation and
$M_R$ for epidemics, as we will explain below.  In a SIR realization,
only one infected cluster emerges for any value of $\beta$.  In
contrast, in a percolation process when $p < 1$ many clusters with a
cluster size distribution are generated \cite{Mey_02}. Thus we need
criteria to distinguish between epidemics (the giant connected cluster
in percolation) and outbreaks (finite clusters). The cluster size
distribution over many realizations of the SIR process, close to but
above criticality, has a gap between small clusters (no epidemics) and
large clusters (epidemics).  Thus when defining a cutoff $s_c$ of the
cluster size as the minimum value before the gap interval, the cluster
sizes below $s_c$ are not considered epidemics but those above $s_c$ are
(see Fig.~\ref{sc11}a).  Note that $s_{c}$ depends on $N$.  Then
averaging those SIR realizations larger than the cutoff $s_c$ we find
that the fraction of recovered individuals $M_R$ maps exactly with $g$
(see Fig.~\ref{sc11}b). In our simulations, we use $s_c=200$ for
$N=10^5$.

When we use a cutoff close to criticality, all the exponents that
characterize the transition are the same for link percolation and
epidemic spreading \cite{Lagorioa2009, Wu_01,Han_01}. Above but close to
$\beta_c^{LP}$ we have
\begin{eqnarray}
M_R(\beta) &\sim& (\beta-\beta_c^{LP})^\alpha,\\
g &\sim& (p-p_c)^\alpha,
\end{eqnarray}
with \cite{Coh_02}
\begin{eqnarray}\label{eq.betat}
 \alpha = \left\{
\begin{array}{ll}1 & \mbox{ for SF with $\nu \geq 4$ and ER networks,}\\
\frac{1}{\nu-3} & \mbox{ for $ 3 < \nu < 4$,}
\end{array}
\right. .
\end{eqnarray}
The exponent $\tau$ of the finite cluster size distribution in
percolation close to criticality is given by
\begin{eqnarray}\label{eq.tau}
 \tau = \left\{
\begin{array}{ll} \frac{5}{2} & \mbox{ for SF with $\nu \geq 4$ and ER networks;}\\
\frac{1}{\nu-2}+2 & \mbox{ for $ 2 < \nu < 4$.}
\end{array}
\right.
\end{eqnarray}

Near criticality the probability of a cluster of size $s$, $P(s)$, has
an exponent $\tau -1$ in which $\tau$ is given by Eq.~(\ref{eq.tau})
(see Fig.~\ref{sc11}a). For SF networks with $\nu \leq 3$ in the
thermodynamic limit, the threshold is zero and there is no percolation
phase transition. In addition, all the exponents take mean-field
values when $\nu\geq 4$ for SF networks and ER networks.

In an uncorrelated network with a degree distribution $P(k)$, the
probability of reaching a node with a degree $k$ by following a randomly
chosen link on the graph is $k P(k) /\langle k \rangle$, where $\langle
k \rangle$ is the average degree. This is because the probability of
reaching a given node by following a randomly chosen link is
proportional to the number of links $k$ connected to that node, and
$\langle k \rangle$ is needed for normalization.  Note that when we
arrive to a node with a degree $k$ by following a random chosen link,
the total number of outgoing links or branches of that node is
$k-1$. Thus the probability of arriving at a node with $k-1$ outgoing
branches by following a randomly chosen link is also $k P(k)/\langle k
\rangle$. This value is the excess degree probability
\cite{New_07,Bra_01}.

To obtain the threshold of link percolation, we consider a
randomly-chosen occupied link. To compute the probability that through
this link an infinite cluster will not be reached we assume, for
simplicity, that we have a Cayley tree with a given degree
distribution. Note that link percolation can be thought of as many
realizations of a Cayley tree with an occupancy probability $p$ that
gives rise to many clusters. The probability that when starting from an
occupied link we will not reach shell $n$ through a path of occupied
links is given by
\begin{eqnarray}
Q_n(p) &=& \sum_{k=1}^\infty \frac{k\;P(k)}{\langle k \rangle}
\left[(1-p)+pQ_{n-1}(p)\right]^{k-1},\\ &=& G_{1}[(1-p)+pQ_{n-1}(p)],
\end{eqnarray}
where $G_{1}(x)=\sum_{k=1}^{\infty}kP(k)/\langle k \rangle x^{k-1}$ is
the generating function of the excess degree distribution. As $n$
increases, $Q_{n}\approx Q_{n-1} = u$ and the probability that we
will not reach an infinite cluster is
\begin{eqnarray}\label{Qinf}
u &=& G_{1}[(1-p)+p \;u].
\end{eqnarray}
Thus the probability that the starting link connects to an infinite
cluster is $f_{\infty}(p)=1-u$. From Eq.~(\ref{Qinf}), $f_{\infty}(p)$
is given by
\begin{eqnarray}\label{finfeq}
f_{\infty}(p)&=&1- G_{1}[1-p\;f_{\infty}(p)].
\end{eqnarray}
The solution to Eq.~(\ref{finfeq}) can be geometrically understood as
the intersection of the identity line $y=x$ and $y=1-G_{1}(1-p\;x)$,
which has at least one solution at the origin, $x=f_\infty(p)=0$, for
any value of $p$. If the derivative of the right hand side of
Eq.~(\ref{finfeq}) with respect to $x$ is
$\left[1-G_1(1-px)\right]'\vert_{x=0}=pG_1'(1)>1$, we will have another
solution in $0<x\leq 1$. This solution $x=f_{\infty}(p)$ is the
probability that a randomly-selected occupied link is connected to an
infinite cluster for a given value of $p$. The criticality corresponds
to the value of $p=p_{c}$ at which the curve $1-G_{1}(1-px)$ has a slope
equal to one. Thus $p_c$ is given by \cite{Coh_01}
\begin{equation}
p_c= {1\over G_1'(1)}={\langle k\rangle\over \langle k^2\rangle -
  \langle k \rangle},
\label{eq:pc}
\end{equation}
which is the same epidemic threshold as that obtained using the SDMP
approach [see Eq.~(\ref{hh_thereshold})].  On the other hand we can
obtain the order parameter of link percolation $g$, which represents the
fraction of nodes that belongs to the giant cluster when a fraction $p$
of links are occupied. The probability that a node with degree $k$ does
not belong to the giant component is given by the probability that none
of its links connect the node to the giant connected cluster (GCC), i.e.,
$\left[1-p\;f_{\infty}(p)\right]^k$. Thus the fraction of nodes that
belong to the GCC is
$g=1-\sum_{k=0}^{\infty}P(k)\left[1-p\;f_{\infty}(p)\right]^k$. Since
the relative epidemic size in the SIR model maps exactly with the
relative size of the giant connected cluster, we find that
\begin{equation}
M_R=g=1-G_0\left[1-p f_\infty(p)\right],
\end{equation}
where $G_{0}(x)=\sum_{k=0}^{\infty}P(k)x^{k}$ is the generating function
of the degree distribution and $ f_\infty(p)$ is the non-trivial
solution to Eq.~(\ref{finfeq}) for $p>p_{c}$. It is straightforward to
show that in ER networks $G_0(x)=G_1(x)=\exp{\left[-\langle k \rangle
    (1-x)\right]}$ and thus $f_{\infty}(p)=M_R$. In pure SF networks
with $1\leq k < \infty$ the generating function of the excess degree
distribution is proportional to the poly-logarithm function
$G_1(x)=Li_{\lambda}(x)/\xi(\lambda)$, in which $\xi(\lambda)$ is the
Riemann function \cite{Bra_01}.

 The LP approach assumes that
    every link does not connect to the GCC with the same probability
    $u$. Note that when we calculate $u$ only the outgoing branches are
    considered in Eq.~(\ref{Qinf}) and some dynamical correlations are
    thus captured. In addition, the LP approach uses the degree
    distribution to describe the network topology. Thus the LP approach
    can predict the final epidemic size and threshold on networks with
    an infinite uncorrelated local tree-like
    configuration~\cite{Newman2001}. Note that SIR spreading is a
    dynamical infection process with an intricate interplay between
    complex network structure and dynamical correlations, and this
    differs from the static link percolation model. Thus the network
    topology, the time evolutions, and the dynamical correlations among
    the states of neighbors cannot be described using the classical LP
    approach, especially near the critical point~\cite{Kenah2007}. For
    epidemic spreading on finite size networks or when there is a
    non-uniform infectious time distribution, research has shown that
    the final state of the SIR model differs from that of the link
    percolation model, including in particular the epidemic threshold,
    mean epidemic size, and epidemic size
    distribution~\cite{Kenah2007,Lagorioa2009}.  A number of advanced LP
    approaches have been developed to address specific
    configurations. For example, Miller et al.~\cite{Miller2007} and
Allard et al.~\cite{Allard2009} generalized the LP approach to study
nodes with different levels of infection and susceptibility, No\"{e}l et
al. developed the LP approach for finite uncorrelated tree-like networks
\cite{Noel2009}, and Marder generalized the LP approach to obtain the
time evolution of the SIR model \cite{Marder2007}. Recently some
researchers also developed LP approaches to address the effects of
clustering \cite{Serrano2007,Newman2009}, degree-degree correlations
\cite{Goltsev2008}, community structure \cite{MIN2012}, and multiplexity
\cite{Dickison2013,Buono2014} in the SIR model. In addition, Newman
derived the LP approach for multiple epidemic spreading dynamics,
analyzed the interactions between the epidemics
\cite{Newman2005,Newman2010,Newman2014}, and found a co-infection
condition in the interacting epidemics. Parshani et al. developed a
modified LP approach to predicting the threshold of the SIS model
\cite{Parshani2010}.

{\bf Message passing (MP) approach.}  In the LP approach the probability of
reaching the infinite cluster by following a randomly chosen link on the
graph is assumed to be the same for all links. This assumption is true
for uncorrelated tree-like networks but is not valid for real-world
networks, which may have, e.g., degree correlation, clustering, and
communities. To take these into account, Karrer and Newman developed the
message passing (MP) approach \cite{Karrer2014} to study the final state of
the SIR model, which differs from the description supplied above. They
assumed that $z_{j\rightarrow i}$ is the probability that node $j$ is
infected by its neighbors in the absence of node $i$, i.e., node $i$ is
in the cavity state. When there is a vanishing small fraction of
initially infected nodes, $z_{j\rightarrow i}$ satisfies the
relationship
\begin{equation} \label{dmp_dis}
z_{j\rightarrow i}=1-\prod_{\ell\in \mathcal{N}(j)\backslash i}
(1-\beta z_{\ell\rightarrow j}),
\end{equation}
where $\prod_{\ell\in \mathcal{N}(j)\backslash i} (1-\beta
z_{\ell\rightarrow j})$ is the probability that node $j$ is not infected
by any neighbors in the absence of node $i$.  Node $i$ is infected by
the epidemic with a probability
\begin{equation} \label{final_dmp_dis}
f_i=1-\prod_{j\in \mathcal{N}(i)}(1-\beta z_{j\rightarrow i}).
\end{equation}
Thus the relative epidemic size is given by
\begin{equation} \label{size}
g=\frac{1}{N}\sum_{i=1}^N f_i.
\end{equation}
To obtain the value of $z_{j\rightarrow i}$ we iterate
Eq.~(\ref{dmp_dis}) from a random initial value and substitute the
results into Eq.~(\ref{final_dmp_dis}) and, using Eq.~(\ref{size}), we
obtain the relative epidemic size of the SIR model.  From
Eq.~(\ref{dmp_dis}),
\begin{equation} \label{edge_rw}
\mathrm{ln}(1+z_{j\rightarrow i})=
\sum_\ell A_{j\ell}\mathrm{ln}(1-\beta z_{j\rightarrow \ell})-
A_{ji}\mathrm{ln}(1-z_{j\rightarrow i}).
\end{equation}
Defining the vectors $\overrightarrow{u}$ and $\overrightarrow{v}$ whose
($j\rightarrow i$)-th components are $u_{j\rightarrow i}
=\mathrm{ln}(1+z_{j\rightarrow i})$ and $v_{j\rightarrow
  i}=\mathrm{ln}(1-\beta z_{j\rightarrow i})$, respectively,
Eq.~(\ref{edge_rw}) can be written
\begin{equation} \label{edge_vec}
\overrightarrow{u}=\textbf{B}\overrightarrow{v},
\end{equation}
where $\textbf{B}$ is the non-backtracking matrix of the network.  If
a global epidemic breaks out, Eq.~(\ref{edge_vec}) will have a
nontrivial solution. Thus the epidemic threshold is the inverse of the
largest eigenvalue of matrix $\textbf{B}$, which is the same as that
described in Eq.~(\ref{threshold_dmp}) for the SIS model predicted by
the DMP approach.

 The advantages and drawbacks of
    the MP approach are the same as those in the DMP approach (see
    details above). Unlike the DMP approach, which describes the time
    evolution of the spreading dynamics, the MP approach uses different
    formulas and considers only the final state of the SIR spreading
    dynamics. Since the MP approach uses a non-backtracking matrix
that allows a description of the full structure of the network but
disallows nodes in the `cavity' state to transmit the epidemic, it
accurately predicts the epidemic size and threshold in artificial and in
some real-world networks~\cite{Wang2015c}.  Recently the MP approach has
been used to control the spread of an epidemic \cite{Altarelli2014}, to
identify patient zero \cite{Antulov-Fantulin2015,Altarelli2013}, and to
locate the most influential seeds \cite{Morone2015,Hu2015}. Using the MP
approach, Morone and Makse studied influence maximization in complex
networks through optimal percolation and found that the low-connected
nodes play an important role in influencing maximization problems
\cite{Morone2015}.  Hu et al.  discovered that the influence
maximization problem is a local optimization problem, not a global one
\cite{Hu2015}.

\begin{table*}
\caption{Some characteristics of the existing approaches used for SIS
  and SIR models including those that take into account network topology
  or describe the dynamical correlations. We indicate when the approach
  is fully ($\checkmark$) or partially ($\clubsuit$) able, or is unable
  ($\times$) to describe the corresponding characteristic. The number of
  equations needed are also listed. Here $n$ is the number of states
  needed to describe each approach. The system size is denoted by $N$
  and $k_{\rm max}$ is the largest degree a node can
  have.} \label{table_Ch} \centering
  \begin{tabular}{|l|c|c|c|c|c|}
    \hline
    \hline
    Approaches & SIS model & SIR model & Network topology & Dynamical correlations
    & Number of needed equations \\
    \hline
    Mean-field (MF)& $\checkmark$ & $\checkmark$ & $\times$  & $\times$ & $1$\\
    Heterogeneous mean-field (HMF)& $\checkmark$ & $\checkmark$ &
    $\clubsuit$ & $\times$ & $k_{\rm max}$\\
    Quench mean-field (QMF) & $\checkmark$ & $\checkmark$ & $\checkmark$
    & $\times$ & $N$\\
    Dynamical message-passing (DMP)  & $\checkmark$ & $\checkmark$ &
    $\checkmark$ & $\clubsuit$ & $N+2E$\\
    Link percolation (LP) & $\clubsuit$& $\checkmark$ & $\clubsuit$ &
    $\clubsuit$& $1$\\
    Edge-based compartmental (EBC)  &$\times$ & $\checkmark$ &
    $\clubsuit$ &$\clubsuit$ & $4$\\
    Pairwise approximation (PA) &$\checkmark$ & $\checkmark$ &
    $\checkmark$ &$\clubsuit$ & $N+E$\\
    Continuous-time Markov (CTM)  &$\checkmark$ & $\checkmark$ &
    $\checkmark$ &$\checkmark$ & $n^N$\\
    \hline
    \hline
  \end{tabular}
\end{table*}

\begin{figure}
\begin{center}
\epsfig{file=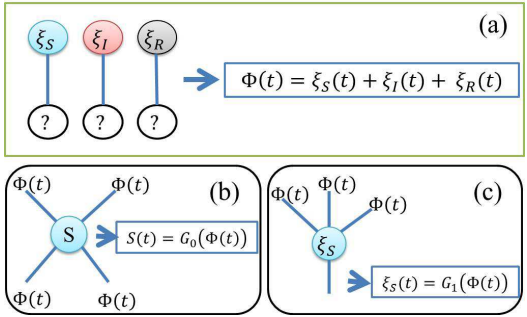,width=0.8\linewidth}
\caption{(Color online) Schematic of
(a) $\Phi(t)$, (b) $S(t)$ and $\xi_S(t)$ of the EBC approach.}
\label{fig1.phi}
\end{center}
\end{figure}

{\bf Edge-based compartmental (EBC) approach.}  Because the LP and MP
approaches are static, they are usually used to address the final state
of the SIR model. To investigate the time evolution of the SIR model,
the edge-based compartmental (EBC) approach was developed
\cite{Miller2011, Miller2012,Volz2008,Volz2011,Valdez2012}. It is based
on the cavity theory (i.e., the MP approach) in which a node $i$ in the
cavity state cannot transmit the infection to its neighbors but can be
infected by its neighbors. Unlike the MP approach in which each edge has
a different probability of transmitting the infection to its neighbors,
the EBC approach makes the same assumption as the LP approach, i.e.,
the probability of infection transmitted
through each link is the same.
The EBC approach is based on a generating
function formalism widely applied to branching and percolation processes
in complex networks. The fraction of susceptible, infected, and
recovered individuals at time $t$ are denoted $s(t)$, $\rho(t)$, and
$r(t)$, respectively.  The EBC approach describes the evolution of the
probability that a denoted root will be susceptible. To compute this
probability, an edge is randomly chosen and a direction given in which a
node $j$ on the target of the arrow is the root, and the base is one of
its neighbors. Disallowing the root $j$ to infect its neighbors,
$\Phi(t)$ is the probability that neighbor $i$ does not transmit the
disease to root $j$, with $\Phi(t)$ given by
\begin{equation}\label{theta}
\Phi(t)=\xi_S(t)+\xi_I(t)+\xi_R(t),
\end{equation}
where $\xi_{S}(t)$, $\xi_{R}(t)$, and $\xi_{I}(t)$ are the probabilities
that the neighbor is susceptible, recovered, or infected but has not yet
transmitted the disease to the root node $j$ [see
  Fig.~\ref{fig1.phi}(a)].  The probability that node $j$ with
connectivity $k$ is susceptible is thus $\Phi(t)^{k}$, and the fraction
of susceptible nodes is given by
\begin{equation}\label{s_t}
s(t)=\sum_{k}P(k)\Phi(t)^k= G_0(\Phi(t)).
\end{equation}
Figure \ref{fig1.phi}(b) shows a schematic of this model.  We next solve
$\xi_S(t)$, $\xi_I(t)$, and $\xi_R(t)$.  A neighbor node $i$ of the root
node $j$ can only be infected by neighbors other than $j$.  Then
node $i$ is susceptible with a probability
\begin{equation} \label{s_s_t}
\xi_S(t)=\frac{\sum_{k}P(k) k\Phi(t)^{k-1}}{\langle
  k\rangle}=G_1(\Phi(t)),
\end{equation}
where $P(k)k/\langle k\rangle$ is the probability that an edge
connects a node with degree $k$ in an uncorrelated network [see
  Fig.~\ref{fig1.phi}(c)]. In the discrete updating method there are
two conditions that allow the increase of $\xi_R(t)$, i.e., (i) the
infected node has not transmitted the infection to $j$ with a
probability $1-\beta$, and (ii) the infected node is removed with a
probability $1$.  Taking these two events into consideration, the
evolution of $\xi_R$ is given by
\begin{equation}\label{xi_R}
\frac{d\xi_R(t)}{dt}=(1-\beta)\xi_I(t).
\end{equation}

At time $t$ the rate of change in the probability that a random edge has
not transmitted the infection is equal to the rate at which the infected
neighbors transmit the infection to their susceptible neighboring nodes
through edges. Thus
\begin{equation}\label{d_theta}
\frac{d\Phi(t)}{dt}=-\beta\xi_I(t).
\end{equation}
Combining Eqs.~(\ref{xi_R}) and (\ref{d_theta}) with initial conditions
$\Phi(0)=1$ and $\xi_R(0)=0$, we obtain
$\xi_R(t)=[1-\Phi(t)](1-\beta)/\beta$, which together with
Eq.~(\ref{d_theta}) and Eq.~(\ref{theta}) allows us to obtain the
evolution of $\Phi(t)$,
\begin{equation}\label{detail_theta_d}
\begin{split}
\frac{d\Phi(t)}{dt} =-\beta \Phi(t)+\beta G_1(\Phi(t))
 +[1-\Phi(t)](1-\beta).
\end{split}
\end{equation}
Using the evolution equations for the infected and removed nodes, which
are $d\rho(t)/dt=-ds(t)/dt- \rho(t)$ and $dr(t)/dt=\rho(t)$,
respectively, we can compute the node density in each state at an
arbitrary time.  In the final state, i.e., $d\Phi(t)/dt=0$,
$\Phi(\infty)=1$ for $t \to \infty$ is a trivial solution of
Eq.~(\ref{detail_theta_d}), and a nontrivial solution emerges only when
$\beta$ is above the critical transmission probability $\beta_c$. Using
an analysis similar to the one used to obtain Eq.~(\ref{h_thereshold}),
the epidemic threshold is
\begin{equation}\label{threshold_eb}
\beta_c^{\rm EBC}=\frac{\langle k\rangle}{\langle k^2\rangle
-\langle k\rangle}.
\end{equation}

In the continuous updating method Eq.~(\ref{xi_R}) is rewritten
\cite{Shu2016}
\begin{equation}\label{xi_R_C}
\frac{d\xi_R(t)}{dt}= \xi_I(t).
\end{equation}
Thus we have $\xi_R(t)=[1-\Phi(t)]/\beta$ and
obtain the epidemic threshold
\begin{equation}\label{threshold_eb_c}
\beta_c^{\rm EBC}=\frac{\langle k\rangle}{\langle k^2\rangle
-2\langle k\rangle}.
\end{equation}

Unlike the LP and MP approaches,
    the EBC approach takes the time evolutions of SIR spreading into
    consideration.  Although the EBC approach also uses the degree
    distribution as the only input parameter to describe network
    topology, it more accurately predicts the epidemic size and
    threshold than the HMF approach. The EBC approach is based on the
    cavity theory in which a node in the cavity state cannot transmit
    infection to its neighbors but can be infected by its
    neighbors. Thus the EBC approach can capture some of the dynamical
    correlations among the states of neighbors.
Researchers have found that the EBC
approach is exact for the SIR model on infinite uncorrelated local
tree-like networks
\cite{Valdez2013,Valdez2013a,Valdez2012a,Miller2014,Wang2014a} not only
in reproducing the dynamics but also
in determining the final state of the model. For example, Wang et
al. generalized the EBC approach to study epidemic spreading on weighted
networks, and found that increasing the heterogeneity of the weight
distribution decreases the size of the epidemic and increases the
threshold \cite{Wang2014a}. Recently these same authors developed the
EBC approach for a non-Markovian social contagion and found a transition
in which the final adoption size depends on such key parameters as the
transmission probability, which can change from discontinuous to
continuous \cite{Wang2015}. The transition can be triggered by such
parameters and structural perturbations to the system as decreasing the
adoption threshold of individuals, decreasing the heterogeneity of the
adoption threshold, increasing the initial seed size or contact
capacity, or enhancing network heterogeneity \cite{Wang2015,Wang2015a,
  Wang2015b}.

\section{Discussion and outlook} \label{sec:dis}

\begin{figure}
\begin{center}
\epsfig{file=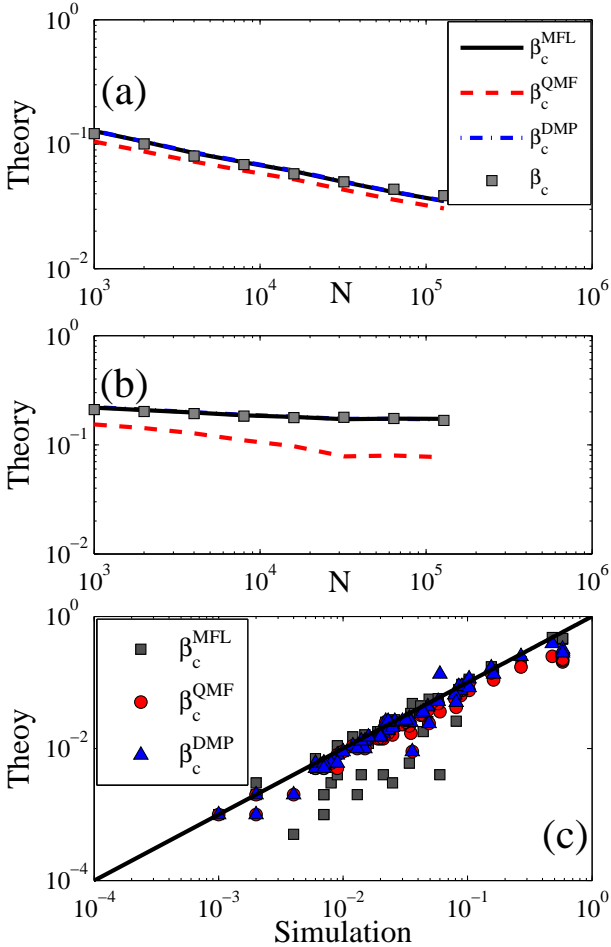,width=1\linewidth}
\caption{(Color online) {\bf Predicting the epidemic threshold for the
    SIR model on uncorrelated networks and $56$ real-world networks.}
  Theoretical predictions of $\beta_c^{\rm MF}$ (black solid lines),
  $\beta_c^{\rm QMF}$ (red dished lines), $\beta_c^{\rm DMP}$ (blue
  dish-dotted lines) and numerical prediction (gray squares) versus
  network size $N$ for power-law degree distribution $P(k)\sim k^{-\nu}$
  with degree exponent $\nu=2.1$ (a) and $\nu=3.5$ (b). In (c), each
  symbol a threshold of a real-world network. $\beta_c^{\rm MFL}$,
  $\beta_c^{\rm QMF}$ and $\beta_c^{\rm DMP}$ are the theoretical
  predictions by the MFL, QMF and DMP, respectively. The value of
  $\beta_c$ is the numerical prediction by using a variability measure
  $\Delta=\sqrt{\langle r^2\rangle -\langle r\rangle^2}/\langle
  r\rangle$ \cite{Shu2015}.}
\label{fig4}
\end{center}
\end{figure}

We have illustrated seven widely used approaches to the dynamics of
epidemic spreading, including the MF, HMF, QMF, DMP, LP, EBC, and
PA. Other not widely used approaches, such as master equations, are
described in Refs.~\cite{Mason2016, Gleeson2011,Gleeson2013}.
Table~\ref{table_Ch} shows which characteristic behaviors are described
by each approach. Note that all seven approaches can be used in the
irreversible SIR model, but that prior to now the EBC has not been used
in the reversible SIS model. Unfortunately none of these approaches can
adequately describe both the full topology of a network and its dynamic
correlations. Table~\ref{table_Ch} shows that in order to more
accurately capture both network topology and dynamic correlations, the
number of required equations increases and they become increasingly
complex. To describe network
    topology, we use the adjacency matrix and non-backtracking matrix as
    illustrated in the QMF and DMP approaches, respectively. To capture
    the dynamical correlations, we use cavity theory to prevent the
    `echo chamber' effect or the evolution of pair node states.

To capture both network topology and dynamical correlations, we adopt
the continuous-time Markov (CTM) approach
\cite{Simon2011,Mieghem2012,Sahneh2013} and find exact results for
epidemic spreading. The CTM approach uses the adjacency matrix to
describe the network topology, and uses the transform matrix generator
$Q_{q^N\times q^N}$ to describe the evolution of the epidemic spreading
and the dynamic correlations.  Once the value of $Q$ is obtained the
probability that a node will be in each state can be computed. Although
the CTM approach provides an exact description it is not widely used in
the field of spreading dynamics because the generator $Q_{q^N\times
  q^N}$ is difficult to obtain, and also it is difficult to solve the
complicated equations, especially for large scale networks. The CTM can
be used, however, to obtain exact solutions in a few specific scenarios
of the SIS model \cite{Sahneh2013}.

For a given epidemic spreading dynamics, these theoretical approaches
sometimes yield different epidemic sizes and thresholds
\cite{Gleeson2008,Li2012a}.  Wang et al. classified the theoretical
approaches into three categories according to the topological information
used~\cite{Wang2015c}. The first is the mean-field like (MFL) approach,
which uses the degree distribution as the sole input parameter. This
category includes the HMF, the LP, EBC, and PA approaches. The second
type is the quenched mean-field (QMF) approach, which describes the
topology of each network using the adjacency matrix. Examples include
the discrete-time Markov chain and the \emph{N}-intertwined approach
\cite{Mieghem2011,Gomez2010}.  The third type is the dynamic message
passing (DMP) approach, which describes network topology in terms of the
non-backtracking matrix. Wang et al. determined the effectiveness of
these three approaches using extensive numerical simulations of the SIR
model on artificial and real-world networks \cite{Wang2015c}.  For
configuration networks they found that the MFL and the DMP approaches
perform better than the QMF approach [see Figs.~\ref{fig4}(a) and
  \ref{fig4}(b)]. For real-world networks, the DMP approach performs
well in most situations [see Figs.~\ref{fig4}(c)].

In summary, we began by describing in terms of increasing complexity
the seven most popular theoretical approaches. We explain their main
ideas and basic assumptions, and we describe the relationships among
them. These approaches
have also been widely used in studying the
dynamics of social contagions
\cite{Dodds2009,Dodds2004,Wang2015,Wang2015a,Wang2015b,
  Huang2016,Majdandzic2016,Zhu2016}. As network science has developed
and expanded, many of the existing theoretical approaches have been
challenged, and we now must take into consideration numerous intricate
mechanisms and network topologies when we build epidemic spreading
models.

The first challenge is how to describe epidemic spreading on complex
real-world networks that are, for example, multilayer or temporal. With
the availability of real-world data, many researchers believe that
treating real-world networks as single or static networks is no longer a
viable approach, and that one must utilize multilayer and temporal
networks \cite{Boccaletti2014,Holme2012}. Although we may want to adopt
the tensor formalism to describe the topology of multilayer and temporal
networks, the network topology and dynamic correlations are difficult to
capture \cite{Arruda2016,Cozzo2013,Domenico2013} with this formalism. In
addition, different spreading mechanisms such as preference spreading
\cite{Xu2015} and layer-switching \cite{Min2016} are produced, and the
intricate network topology further increases the difficulty of
describing the spreading of an epidemic.

The second challenge is how to describe epidemic spreading once human
behavior is included. Human behavior will markedly affect epidemic
spreading dynamics \cite{Mieghem2013,Jo2014} due to burst, memory
\cite{Min2011,Zhou2012,Barabasi2005, Song2010}, and mobility effects
\cite{WangLin2014,WangLin2013,Tang2009,wang2015coupled,
  zhang2016impact,zhang2013braess}.  These features induce a
non-Markovian effect in the spreading dynamics that causes strong
dynamic correlations that are difficult to describe
\cite{Altarelli2014}. The existing theoretical approaches can address
only some specific situations, and a general framework for non-Markovian
spreading dynamics is still lacking.

A third challenge is how to describe coevolution spreading dynamics. In
real-world systems when two strains of the same disease spread in the
same population and interact through cross-immunity \cite{Karrer2010,
  Sanz2014,Marceau2012} or mutual reinforcement \cite{Cai2015}, the
information from each competes for the limited attention-span of the
participants \cite{Gleeson2014}, and there is an asymmetric interaction
between the spread of information and the spread of the epidemic
\cite{Wang2016,Wang2014,Liu2015a,Granell2013}.  An accurate, unified
theoretical approach for coevolution dynamics is still lacking and
presents great challenges because, in this case, the dynamic
correlations are enhanced.

\acknowledgments

This work was partially supported by the National Natural Science
Foundation of China under Grant Nos.~11105025, 11575041, 61433014 and
China Scholarship Council.  The Boston University work was supported
by DTRA Grant HDTRA1-14-1-0017, by DOE Contract DE-AC07-05Id14517, and
by NSF Grants CMMI 1125290, PHY 1505000, and CHE-1213217. LAB
knowledge the support of UNMdP and FONCyT, PICT 0429/13.

\section*{Definitions and abbreviations of parameters}

\begin{table*}
\caption{Definitions of parameters and abbreviations.} \label{table1}
  \centering
  \begin{tabular}{|c|l|}
    \hline
    \hline
   Parameter/Abbreviation & Definition \\
   \hline
    MF approach& Mean-field approach\\
    \hline
    HMF approach& Heterogeneous mean-field approach\\
   \hline
    QMF approach& Quench mean-field approach\\
    \hline
   DMP approach& Dynamical message-passing approach\\
    \hline
    PA approach&  Pairwise approximation approach\\
    \hline
    LP approach & Link percolation approach\\
    \hline
    EBC approach & Edge-based compartmental approach\\
    \hline
    CTM approach & Continuous-time Markov approach\\
    \hline
    GCC & Giant connected cluster\\
    \hline
    $P(k)$ & Degree distribution \\
    \hline
    ${\bf A}$ & Adjacency matrix \\
    \hline
    $\textbf{B}$ & non-backtracking matrix \\
    \hline
    $E$ & Number of edges in the network \\
    \hline
    $N$ & Network size \\
    \hline
    $p$  & Edge occupancy probability \\
    \hline
    $g$ & The relative size of the GCC\\
    \hline
    $u$ & The endpoint of a randomly selected edge is not connected to the GCC\\
    \hline
    $t_r$ & Recovery time\\
    \hline
    $f_i$ & Probability that node $i$ is infected by neighbors\\
    \hline
    $\mathcal{N}(j)$ & Neighbor set of node $j$\\
    \hline
    $G_0(x)$ & Generation function of the degree distribution\\
    \hline
    $G_1(x)$ & Generation function of the excess degree distribution\\
    \hline
    $M_R$ & The final fraction of recovered nodes\\
    \hline
    $\nu$ & Exponent of power-law degree distribution \\
    \hline
    $\langle k\rangle$ & Average degree \\
    \hline
    $\langle k^2\rangle$ & Second moment of the degree distribution \\
    \hline
    $k_{\rm max}$ &  Maximum degree of networks\\
    \hline
    $\lambda$ & Infection transmission rate \\
    \hline
    $\gamma$ & Recovery rate\\
    \hline
    $\beta$ & Effective transmission rate\\
    \hline
    $\rho(t)$ & Fraction of infected node at time $t$\\
    \hline
    $\overrightarrow{\rho}(t)$ & Vector of $\rho_k(t)$ \\
    \hline
    $\rho(\infty)$ & Fraction of infected node in the steady state\\
    \hline
   $s(t)$ & Fraction of susceptible node at time $t$\\
   \hline
   $s_c$ & Cutoff in the cluster size to split epidemic from outbreaks\\
    \hline
    $\rho_k(t)$ & At time $t$, the fraction of infected node with degree $k$\\
    \hline
    $s_k(t)$ & At time $t$, the fraction of susceptible node with degree $k$\\
    \hline
    $\Theta(t)$ & Probability that a susceptible node connects to an infected neighbor \\
    \hline
    $\rho_i(t)$ & Probability of node $i$ in the infected state at time $t$\\
    \hline
    $s_i(t)$ & Probability of node $i$ in the susceptible state at time $t$\\
    \hline
    $r(t)$ & Fraction of removed node at time $t$\\
    \hline
    $\beta_c$ & Epidemic threshold\\
    \hline
    $p_c$ & Critical edge occupied probability\\
    \hline
    $\beta_c^{\rm MF}$ & Epidemic threshold predicted by the MF approach\\
    \hline
    $\beta_c^{\rm HMF}$ & Epidemic threshold predicted by the HMF approach\\
    \hline
    $\beta_c^{\rm QMF}$ & Epidemic threshold predicted by the QMF approach\\
    \hline
    $\beta_c^{\rm DMP}$ & Epidemic threshold predicted by the DMP approach\\
    \hline
    $\beta_c^{\rm SQMF}$ & Epidemic threshold predicted by the simplified QMF approach\\
    \hline
    $\beta_c^{LP}$ & Epidemic threshold predicted by the LP approach\\
    \hline
    $\beta_c^{\rm EBC}$ & Epidemic threshold predicted by the EBC approach\\
    \hline
    $ C $ & Jacobian matrix of $ C _{kk^\prime}=\beta
    kk^\prime P(k^\prime)/\langle k\rangle-\delta_{k,k^\prime}$\\
    \hline
    $\delta_{k,k^\prime}$ & Dirac delta function \\
    \hline
    $\Lambda_A$  & Largest eigenvalue of adjacency matrix \\
    \hline
    $\Lambda_\textbf{B}$  & Largest eigenvalue of non-backtracking matrix \\
    \hline
    $\overrightarrow{f}(\Lambda_A)$ & Eigenvector of $\Lambda_A$\\
    \hline
    $f_i(\Lambda_A)$ & The $i$-th element of the eigenvector
    $\overrightarrow{f}(\Lambda_A)$ of $\Lambda_A$\\
    \hline
    $f_{\infty}(p)$ & Probability that the a random chosen link connects to the GCC\\
    \hline
    $f_i$ & Probability that node $i$ connects to the GCC\\
    \hline
    $\theta_{j\rightarrow i}(t)$ & Probability that node $j$ is infected
    by neighbors at time $t$ in the absence of neighbor $i$\\
    \hline
    $z_{j\rightarrow i}$ & Probability that node $j$ is infected
    by neighbors in the absence of neighbor $i$\\
    \hline
    $\psi_{x_i x_j}(t)$ & Probability
that nodes $i$ and $j$ are in the $x_i$ and $x_j$ state, respectively\\
    \hline
    $\phi_{x_ix_jx_\ell}(t)$ & The probability that nodes $i$, $j$
    and $\ell$ are in the $x_i$, $x_j$ and $x_\ell$ state, respectively \\
    \hline
    $\Phi(t)$ & Probability that a randomly selected edge has not transmitted the
infection to a neighbor by time $t$\\
    \hline
    $\xi_S(t)$ & Probability that a neighbor of a node is in the susceptible state
    at time $t$\\
    \hline
    $\xi_I(t)$ & Probability that a neighbor of a node which is in the infected state
    and has not transmitted the infection to it by time $t$ \\
    \hline
    $\xi_R(t)$ & Probability that a neighbor of a node is in the recovered state
    and has not transmitted the infection to it by time $t$ \\
    \hline
    \hline
  \end{tabular}
\end{table*}


\end{document}